\documentclass[a4paper,fleqn]{aa}
\usepackage{inputenc}
\usepackage{amsmath, amssymb}
\usepackage{graphicx}
\usepackage{subcaption}
\usepackage{booktabs}
\usepackage{natbib}
\bibpunct{(}{)}{;}{a}{}{,} 
\usepackage{hyperref}

\newcounter{refcount}

\newcommand{\Ncri}{\ensuremath{N_{\rm{critical}}}}
\newcommand{\Nbg}{\ensuremath{N_{\rm{bg}}}}
\newcommand{\Nfg}{\ensuremath{N_{\rm{fg}}}}
\newcommand{\sigmabg}{\ensuremath{\sigma_{\rm{bg}}}}
\newcommand{\sigmafg}{\ensuremath{\sigma_{\rm{fg}}}}
\newcommand{\Nobs}{\ensuremath{N_{\rm{obs}}}}

\newcommand{\uncert}{\ensuremath{\tilde{\sigma}_S}}

\newcommand{\myimage}[1]{\begin{center}\includegraphics[angle=0, width=0.9\textwidth]{images/#1}\end{center}}

\newcommand{\myimagesmall}[1]{\begin{center}\includegraphics[angle=0, width=0.48\textwidth]{images/#1}\end{center}}
\newcommand{\myimages}[2]{\begin{center}\includegraphics[angle=0, width=#2\textwidth]{images/#1}\end{center}}

\newcommand{\myimageTwo}[2]{\begin{center}\includegraphics[angle=0, width=0.45\textwidth]{images/#1}
    \includegraphics[angle=0, width=0.45\textwidth]{images/#2}\end{center}}

\addto\extrasenglish{%
\let\subsectionautorefname\sectionautorefname

}

\makeatletter
\renewcommand*\aa@pageof{, page \thepage{} of \pageref*{LastPage}}
\makeatother

\title{Selection functions of large spectroscopic surveys}
\titlerunning{Selection function.}
\author{Alexey Mints\inst{\ref{inst1},~\ref{inst2}}\thanks{email: mints@mps.mpg.de} \and Saskia Hekker\inst{\ref{inst1},~\ref{inst2}}}
\authorrunning{A. Mints and S. Hekker}
\institute{Max Planck Institute for Solar System Research, Justus-von-Liebig-Weg 3, 37077 Göttingen, Germany \label{inst1} \and Stellar Astrophysics Centre, Department of Physics and Astronomy, Aarhus University, Ny Munkegade 120, DK-8000 Aarhus C, Denmark \label{inst2}}

\abstract
{Large spectroscopic surveys open the way to explore our Galaxy. In order to use the data from these surveys to understand the Galactic stellar population, we need to be sure that stars contained in a survey are a representative subset of the underlying population. Without the selection function taken into account, the results might reflect the properties of the selection function rather than those of the underlying stellar population.} 
{In this work, we introduce a method to estimate the selection function for a given spectroscopic survey. We apply this method to a large sample of public spectroscopic surveys.} 
{We apply a median division binning algorithm to bin observed stars in the colour-magnitude space. This approach produces lower uncertainties and lower biases of the selection function estimate as compared to traditionally used 2D-histograms. We run a set of simulations to verify the method and calibrate the one free parameter it contains. These simulations allow us to test the precision and accuracy of the method.}
{We produce and publish estimated values and uncertainties of selection functions for a large sample of public spectroscopic surveys. We publicly release the code used to produce the selection function estimates.}
{The effect of the selection function on distance modulus and metallicity distributions of stars in surveys is important for surveys with small and largely inhomogeneous spatial coverage. For surveys with contiguous spatial coverage the effect of the selection function is almost negligible.}

\keywords{Stars: distances -- Stars: fundamental parameters -- Galaxy: stellar content}

\date{XXX/YYY}

\begin{document}

\renewcommand{\figureautorefname}{Fig.} 
\renewcommand{\sectionautorefname}{Section} 
\renewcommand{\subsectionautorefname}{Section} 

\maketitle
\section{Introduction}\label{sec:intro}
Large stellar spectroscopic surveys aim at probing the stellar population properties throughout the Galaxy. With the aid of modern technology it is possible to perform spectroscopic surveys observing millions of stars.
Depending on the goals and the used instrument, surveys can differ in depth, spatial coverage and can select different kinds of stars for observations. Moreover, it is feasible to observe only a tiny fraction of all stars that are observable in our Galaxy, even if we consider only stars bright enough to be observed with modern instruments. To probe the underlying stellar populations we have to know what fraction of stars was observed, in order to correct for possible selection biases or to prove an absence thereof.

There are several possible questions we might want to answer, regarding a given spectroscopic survey:
\begin{enumerate}
    \item What is the fraction of stars in the footprint of each plate or field of view in a survey that was observed compared to the number of stars available for observations in the same area;
    \item What is the fraction of stars in the selected area on the sky that was observed compared to the number of stars available for observations in the same area;
    \item What is the fraction of stars in the selected area on the sky that was observed compared to the \textit{total} number of stars in the same area;
\end{enumerate}

The first two questions can be answered by comparing stellar number counts for a spectroscopic survey with stellar number counts for some photometric survey. The photometric survey has to be chosen such that it is complete at least down to the faintest stars in the spectroscopic survey. Best results can be achieved if the target allocation strategy for the spectroscopic survey can be directly converted to the selection function. This, however, is not always possible due to proprietary nature of the photometric survey used and complexity in target allocation strategy. Another difficulty arises from the fact that not for all targets observed within a survey spectroscopic parameters have been measured, due to the limitations of the model spectra grids, low signal-to-noise ratios and other problems.

The derivation of the selection function by comparison of a spectroscopic survey to a photometric one produces useful results only when observed stars are a representative subset of the stellar population at a given area on the sky. This is true when only broad-band photometry was used for the target allocation process, as such photometry is almost insensitive to the population properties. In that case we can assume that the selection function depends exclusively on photometric magnitudes and colours. This should generally hold even if targets were selected from a photometric survey that is different from the one used to estimate the selection function. 
However, this assumption breaks down when additional data are used or if some specific fields are observed. For example, the Gaia-ESO survey \citep{GAIA_ESO} contains a large set of fields that are positioned at open clusters, with possible cluster members selected as spectroscopic targets. For these fields selection functions cannot be reliably estimated by comparing spectroscopic and photometric surveys. This is because cluster members are often selected by means other than photometry, for example, using proper motions and parallaxes.
In that case, we cannot any more assume that stars observed in a given range of magnitudes and colours are representative subsample of all stars in that range.
 Another example for which the assumption that the observed sample of stars is representative subsample of the stellar population breaks, is the APOGEE survey \citep{APOGEE}. There, additional narrow-band photometry was used to select giant stars over main-sequence dwarfs \citep{2013AJ....146...81Z}. Ignoring this fact will lead to erroneous results for the selection function.

Calculation of the selection function on plate-by-plate basis is more straightforward and potentially more precise than doing that for arbitrary sky regions. The reason for that is that in that case we compare the observed sample with the exactly the same photometric set of stars that was used in the target allocation process. So limiting ourselves to the plate area only, we can expect to reconstruct the selection function with higher precision. Another argument for this strategy is that the target allocation strategy could change between plates, even if they cover the same region on the sky. Thus a selection function for a combination of plates might be more complex than that for a single plate. 

On the other hand, dealing with sky areas has its own advantages. First, choosing sky areas that are larger than a single field of the survey can substantially increase the source statistics and with that reduce the uncertainty of the selection function estimate. Second, the choice of sky areas can be advantageous for further analysis (like fitting a galactic model) and comparison of results from different surveys with overlapping footprints. It is also possible to take overlapping plates and repeated observations of same targets into account -- this can be accounted for before the selection function is calculated, and each star will enter the analysis only once no matter how many times it was observed. The drawback of this approach is that observations might cover only a fraction of the selected area, and thus are not representative for the stellar population of this area.
 
In recent works by \citet{2016MNRAS.460.1131S, 2017MNRAS.468.3368W, 2017A&A...606A..97N} and \citet{2018MNRAS.476.3278C}, selection functions for a set of spectroscopic surveys were studied. Using the derived selection function, these authors tested if there are any selection biases in the studied spectroscopic survey. The most common approach is to process a survey in a plate-by-plate manner. 

\citet{2016MNRAS.460.1131S} derived the selection function for a subset of the Gaia-ESO survey, using the available information on target allocation strategy. A number of Gaia-ESO fields is dedicated to open cluster studies and was therefore excluded from the analysis. In \citet{2016MNRAS.460.1131S}, 2MASS and VHS photometry were used to derive the selection function. The number of observed stars was compared to the number of stars in the photometric survey for each $0^m.05 \times 0^m.5$ bin of the colour-magnitude space. Figure 19 in \citet{2016MNRAS.460.1131S} shows that the selection function has a large effect at least for the metallicity distribution function (MDF) of the survey. A table containing selection function values for almost 10,000 stars was published.

\citet{2017MNRAS.468.3368W} studied the selection function of the RAVE survey and its effects on kinematic and chemical biases. Selection functions were calculated both on a plate-by-plate basis and for 5th order HEALPix sky cells (see \autoref{sec:healpix} below). In both cases, the selection function was calculated as a ratio of the number of observed stars in an $I$-band magnitude bin to the number of 2MASS stars in the same bin. The $I$-band magnitude for 2MASS stars was calculated using colour-dependent correction of 2MASS $J$-band photometry. $I$-band magnitude bins with a width of $0^m.1$ were used. On top of the photometry-based selection, a pipeline selection function was calculated to account for stars observed by RAVE for which no stellar parameters were derived.
Using simulations of the RAVE survey with Galaxia \citep{Galaxia}, \citet{2017MNRAS.468.3368W} concluded that the selection function of RAVE survey does not have an effect on observed kinematic and chemical distributions. A table containing data on the selection as a function of $I$-band magnitude is available on the RAVE web page\footnote{\url{https://www.rave-survey.org/downloads}}.

\citet{2017A&A...606A..97N} studied the effect of the selection function on the MDF in APOGEE \citep{APOGEE}, LAMOST \citep{LAMOST}, RAVE and Gaia-ESO surveys. This was done by building a histogram in colour-magnitude space for sources observed in each field and comparing it to a similar histogram for sources from a photometric survey in the same area. Photometry was taken from different sources to match the depth and target allocation strategy of each survey. The bin sizes for the histogram in colour-magnitude space were $0^m.05$ in colour and $0^m.3$ in magnitude for all surveys. \citet{2017A&A...606A..97N} studied the effect of the selection function using Galaxia \citep{Galaxia} and TRILEGAL \citep{TRILEGAL}. The comparison was focused on the MDF for sources in a range of Galactic coordinates with and without the effect of the selection function (see their Figure 10). \citet{2017A&A...606A..97N} concluded that the selection function has almost no effect on the MDF and observed metallicity gradients. They note, however, that the selection function effect is largest for Gaia-ESO survey, which they attribute to the Poisson noise. Moreover, discrepancies are also visible for APOGEE survey. Notably, differences in the MDF between  APOGEE and Gaia-ESO surveys and corresponding models are larger than those for RAVE and LAMOST. Derived values of the selection function were not published.

\citet{2018MNRAS.476.3278C} calculated the selection function for LAMOST Galactic anti-center survey (LAMOST-GAC) data release 2 \citep{LAMOST_GAC_2}. As in \citet{2017A&A...606A..97N}, a ratio of two histograms (one for spectroscopic sources, one for photometric ones) was used to estimate the selection function. Photometry was taken from XSTPS-GAC or APASS \citep{APASS}. Histograms were made in the space of $g-r$ colour and $r$ magnitude with bin sizes of $0^m.25$ in colour and $0^m.2$ in $r$ magnitude. Similarly to \citet{2017MNRAS.468.3368W}, an additional term was added to the estimate of the selection function to accommodate for sources for which no stellar parameters were derived. They also confirm the result of \citet{2017A&A...606A..97N} that the selection has little effect on the MDF for LAMOST. Derived values of the selection function were not published.

Overall, the trend is that the selection function is more important for surveys with less homogeneous sky coverage, like APOGEE and Gaia-ESO, and almost negligible for surveys with contiguous footprint, like RAVE and LAMOST.

Answers to the third question regarding the number of observed stars with respect to the total number of stars, are model dependent and involve assumptions on stellar evolution and stellar luminosity functions. Generally, we need to predict how many faint stars correspond in a given population to a given number of observed brighter stars. This can be done, for example, by modelling the complete stellar population, having the observed age and metallicity distributions and then calculating the fraction of this population that falls into the observed range of colours and magnitudes. In the case of SEGUE it was possible to use a simplified approach \citep{2012ApJ...753..148B}, given that for main sequence stars observed by that survey colours and magnitudes have little dependence on age. In a general case, we need to know the distribution of stars in distance, metallicity and age to estimate the number of unobserved stars. This task is beyond the scope of this study.

The aim of this work is to set up a method of obtaining unbiased estimates of the selection function for an arbitrary survey. We also produce the estimated uncertainties, which are important if we want to analyse the significance of the selection function effect. The derived method is applied to public spectroscopic surveys, including those for which no study on the selection function was published so far (like LAMOST and GALAH).

\section{Photometric selection function}\label{sec:method}
The most basic definition of a selection function is the ratio of the number of spectroscopically observed stars to the total number of stars with similar properties (for example, location on the sky, visible magnitudes and colours). Hence, in order to estimate a selection function we need to bin the data in sky coordinates, visible magnitudes and colours and count the number of spectroscopically observed stars and the total number of stars. In this section, we describe how this division is done and how the selection function is then estimated in each bin.

\subsection{HEALPix grid construction}\label{sec:healpix}
Considering arguments discussed in \autoref{sec:intro}, we chose to calculate the selection function using fixed sky areas that will be the same for all surveys rather than to work with single fields in each survey. To divide the sky into equal area parts, we use the Hierarchical Equal Area iso-Latitude Pixelization (HEALPix) tool \citep{2005ApJ...622..759G}. We used three orders of this pixelization (3, 4 and 5), with areas approximately 53.7, 13.43 and 3.36 square degree. This allows us to find a balance between the number of spectroscopically observed stars in the HEALPix cell and the variations of the background across the sky within that cell. Galactic coordinates were used for the pixelization, as it naturally groups HEALPix sky cells in galactic latitudes, which can be useful in further analysis. 

\subsubsection{Variation of background in HEALPix cell}
\begin{figure}
    \begin{center}
        \myimagesmall{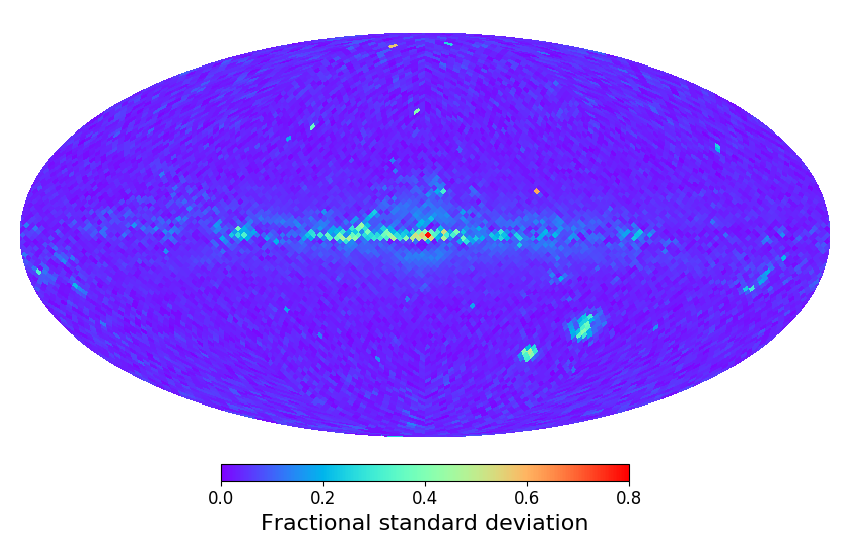}
    \end{center}
    \caption{An illustration of the 2MASS star density variation across the sky. Each 5th order HEALPix cell (see \autoref{sec:healpix}) is colour-coded by the fractional standard deviation of stellar density within that cell.}\label{fig:healpix}
\end{figure}

When we use the colour-magnitude distribution of background stars for a given HEALPix area, we have to be aware of the variations  of the stellar number density within this area. In \autoref{fig:healpix} we give an illustration of an amplitude of this variation for 2MASS sources. In this Figure, each 5th order HEALPix cell $C_5$ is colour-coded by the fractional standard deviation of the stellar number density, calculated from four 6th order cells within $C_5$. These fractional standard deviations are highest around the Galactic centre, Magellanic clouds and large clusters and can be as high as 79 percent. Outside of the galactic disc standard deviations are much smaller -- on the order of one percent or less.

\subsection{Colour-magnitude diagram binning}
\subsubsection{Histogram binning}\label{sec:hess_binning}
\begin{figure}
    \begin{center}
        \myimages{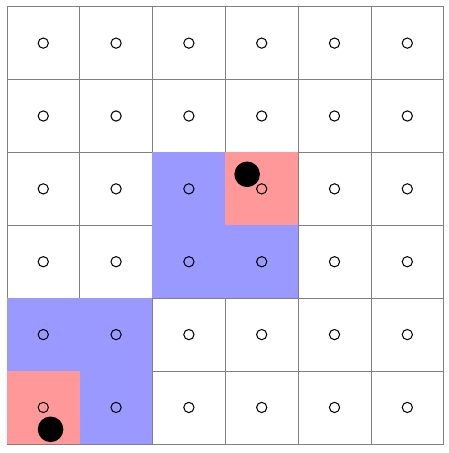}{0.3}
    \end{center}
\caption{Illustration of the effect of binning on the selection function estimate. Black dots illustrate foreground sources, circles are background sources. For small bins (filled with red), $S = 1$, for four times larger bins (filled with blue) $S = 1/4$, while for the even larger bin (full grid) $S = 2 / 36$.}\label{fig:hess_binning}
\end{figure}

The common approach in estimating the selection function is to build histograms of the source distribution in the colour-magnitude space for a selected area on the sky. Then one has to count the number of background (those from the photometric survey) sources $\Nbg$ and the number of foreground sources (for which spectra were obtained) $\Nfg$ in each bin of the histogram. The ratio $S = \Nfg / \Nbg$ will produce an estimate of the selection function in this bin. The main drawback of this approach is that the result depends on the bin size. For larger bins, information about the selection function variation within the bin is lost. In the extreme case of a just one large bin containing all sources, $S$ is equal to the ratio of the number of stars in the spectroscopic survey to the number of stars in the photometric survey. This value is a general property of the spectroscopic survey and cannot be used to infer the effect of the selection on a star-by-star basis.
On the other hand, for smaller bins the statistic can be too low for a reliable estimation. Most importantly, there is a trend to overestimate the selection function, if not all bins are populated with foreground sources. This is illustrated in \autoref{fig:hess_binning}: depending on the chosen bin size, selection function varies by a factor of $18$ between $1$ and $2/36$. The problem is caused by the fact that the selection function is evaluated only at bins where foreground sources are found, which produces a systematic bias in the estimates.

In order to mitigate the above problems, we need to find a way to increase the resolution of the selection function estimate while keeping the number of stars used to derive the value of $S$ in each point above a certain minimum number, which provides lower uncertainty. We therefore introduce the median division binning.

\subsubsection{Median division binning}\label{sec:median_binning}
\begin{figure}
    \myimagesmall{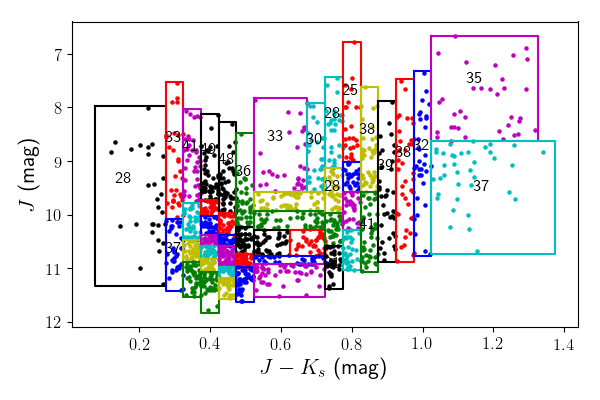}
    \caption{Illustration of the median division algorithm output. Colours are used only to separate cells visually. For this plot, $\Ncri = 25$ was taken. Numbers indicate the number of points in each cell, and are by construction between $\Ncri$ and $2\times\Ncri$. For smaller cells numbers are not shown for visual purpose. }\label{fig:median_binning}
\end{figure}

In order to mitigate problems arising when histograms in colour-magnitude space are used to estimate the selection function, we use a ``median division'' scheme, similar to the one described in \citet{2006MNRAS.373.1293S}  and implemented in the EnBiD code \citep{2011ascl.soft09012S}. We aim at dividing a colour-magnitude plane into rectangular cells (not to be confused with HEALPix sky-cells) in such a way that each cell contains at least $\Ncri$ foreground sources.  
Let $(x_i, y_i)$ be the coordinates of points on the plane. We then consider a rectangular cell with lower left corner at $(x_{min}, y_{min})$ and upper right corner at $(x_{max}, y_{max})$. We calculate the Shannon entropy $H$ along each axis:
\begin{equation}
H = - \sum_i P_i \log P_i,
\end{equation}
where $P_i$ are the values of the histogram build for $x$ or $y$ values. We then select for the next division the axis ($x$ or $y$) for which this entropy is smallest. Let's assume, that the $x$ axis has the lower entropy. We take the median value $x_m = <x_i>$ and divide the cell into two sub-cells for which $x \leq x_m$ and $x > x_m$. This process is repeated recursively for each of the resulting two cells. Recursion stops, when cells contain less than $N_{min} = 2\times \Ncri$ number of points. No further divisions are applied, as these would produce two cells with at least one of them having less than $N_{min} / 2 = \Ncri$ points. Hence each cell contain between $\Ncri$ and $N_{min} = 2\times \Ncri$ points. The result of this process is illustrated in \autoref{fig:median_binning}.

We applied median division binning in the colour-magnitude space for the set of foreground stars in each HEALPix sky-cell. For each colour-magnitude cell produced by median division binning we obtain a number of foreground stars in that cell $\Nfg$ and a number of background star in the same cell $\Nbg$. 
For the background we used the distribution in the $J$ versus $J-K_s$ plane of 2MASS stars from the same HEALPix sky-cell. This was represented as a two-dimensional histogram, with bin size of $0^m.05 \times 0^m.05$. The bin size was chosen to be approximately twice the mean 2MASS photometric uncertainty. Photometric uncertainty will smear out all variations of the selection function $S$ on scales smaller than $0^m.05$, so choosing a smaller bin size will not change our results. Choosing larger bins, however, might cause the loss of information on variations of $S$. Median binning cell borders were forced to align with photometric histogram bin edges. This sets a lower limit on the cell size -- a cell cannot be smaller than one histogram bin. 

The only free parameter of this method is $\Ncri$. In \autoref{sec:tests} below we explore how $\Ncri$ influences the precision and accuracy of our estimates and propose a method of choosing its value. 

\subsection{Uncertainty of selection function estimation}\label{sec:uncertainty}
An important parameter of the selection function estimation is the uncertainty of the result. In this work we produce a formal uncertainty along with the selection function value. It is important to know the uncertainty of the selection function: if we are to build further conclusions on selection-corrected samples, we need to know the uncertainties of the corrected values.

\subsubsection{Counts statistics}
We assume that in the process of target selection sources are randomly picked from $\Nbg$ sources in a given colour-magnitude cell with a probability $S$. 
Hence we can assume that the number of selected targets $\Nfg$ follows a binomial distribution with a mean $\Nbg\times S$ and variance $\Nbg \times S (1 - S)$. 
We use the number of foreground $\Nfg$ and background $\Nbg$ stars to estimate $S$ and its uncertainty $\sigma_S$. Here, we take the estimate of the standard deviation of the binomial distribution for the uncertainty $\sigma_S$. This is done using the method developed by \cite{10.2307/2685469}:
\begin{align}
   S &= \frac{\Nfg + 1/2}{\Nbg + 1} \label{eq:selection} \\
   \sigma_{S} &= \sqrt{\frac{S (1 - S)}{\Nbg + 1}} \label{eq:uncertainty} 
\end{align}

In the limit of large $\Nfg$ and $\Nbg$, as well as $S \ll 1$, constants ($1/2$ and $1$) can be dropped and \autoref{eq:selection} and \ref{eq:uncertainty} turns into:
\begin{align}
S_{approx} &= \frac{\Nfg}{\Nbg}  \label{eq:selection_approx} \\ 
\sigma_{S, approx} &= \frac{\sqrt{\Nfg}}{\Nbg}.  \label{eq:uncertainty_approx} 
\end{align}
So the fractional uncertainty decreases approximately as $\Nfg^{-1/2}$, which is similar to Poisson statistics.

\subsubsection{Background variation within the bin}\label{sec:bg_variation}
Both background and foreground densities can vary substantially within each colour-magnitude cell that we build. This affects the difference between the estimate of the selection function and its true value. We take that into account calculating uncertainties in the following manner. 
For a given median binning cell, we take fore- and background source counts for each photometric histogram bin in that cell.
We then use standard deviations of these counts $\sigmafg$ and $\sigmabg$ as measures of fore- and background source density variations. Hence, instead of \autoref{eq:uncertainty} we use for the uncertainty of the selection function the following expression:
\begin{equation}
  \sigma_S = S \sqrt{\frac{(1 - S)}{S(\Nbg + 1)} + \left(\frac{\sigmafg}{\Nfg}\right)^2 + \left(\frac{\sigmabg}{\Nbg}\right)^2}.\label{eq:uncertainty_corrected}
\end{equation}
The smallest median division binning cell size is just one bin, so $\sigmabg = \sigmafg = 0$, and no correction is added.

\subsection{Improvements for median division binning}\label{sec:improvements}
We introduce several improvements of the median division binning algorithm to increase its precision and reduce bias. These are described below.

\subsubsection{Cell shrinking}\label{sec:shrinking}
\begin{figure}
    \myimagesmall{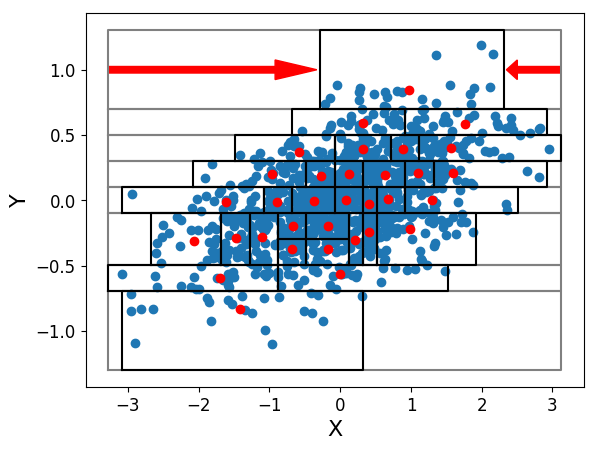}
    \caption{Shrinking and interpolation illustration. Blue points show a random distribution of points to which the median division binning was applied. Grey and black lines show cell borders before and after shrinking is applied, respectively. Red arrows illustrate the direction of shrinking for the uppermost cell. Red points illustrate the placement of interpolation nodes (at the centre of mass of each cell).}\label{fig:shrink}
\end{figure}

We use the assumption that sources at the edges of the area in the colour-magnitude space covered by foreground sources represent real edges beyond which no sources were targeted, and thus $S \equiv 0$ outside this area. 
At the end of the median binning process, outer bins will extend to the edges of the initial distribution of the foreground stars, as illustrated in \autoref{fig:shrink}, which might include areas in the colour-magnitude space that were not included in the survey. To mitigate this problem, we applied a ``shrinking'' procedure, shifting outer borders to the location of the outermost foreground star in each cell. Only outer border were shifted to make sure that the area covered by cells does not contain gaps. 

\subsubsection{Interpolation} \label{sec:interpolation}
In order to further improve the selection function estimate, we performed an interpolation between colour-magnitude cells produced by median binning to obtain values of the selection function and its uncertainty at locations of each spectroscopic source in the colour-magnitude diagram. We used the mean positions of foreground stars in each cell as interpolation nodes, as illustrated in \autoref{fig:shrink}. The interpolation was performed by applying Delaunay triangulation to the set of nodes and fitting a plane through each triangle (simplex), as it is implemented in \textit{SciPy}\footnote{See \texttt{scipy.interpolate.LinearNDInterpolator}, \citet{SciPy}}.
For points outside of the polygon containing all interpolation nodes (convex polygon), extrapolation was used. To extrapolate to a given position on the colour-magnitude diagram, we fit a plane through 10 nodes nearest to that point, and used the value predicted by that plane at a given position. 

We used linear interpolation in the logarithmic scale to obtain values of $\log \Nfg$ and $\log \Nbg$ at the location of each star on the colour-magnitude diagram and then produced estimates of the selection function $S$ and its uncertainty $\sigma_S$ using \autoref{eq:selection} and \ref{eq:uncertainty_corrected}. The  logarithmic scale for the interpolation is beneficial for our task, as it naturally avoids negative values. We have verified that calculating $S$ and $\sigma_S$ in each cell and then interpolating them gives only a marginal difference in the result. We also tested if the interpolation improves the estimate for histogram binning and found only a marginal improvement.

\section{Testing on simulated data}\label{sec:tests}

\begin{figure*}
    \myimage{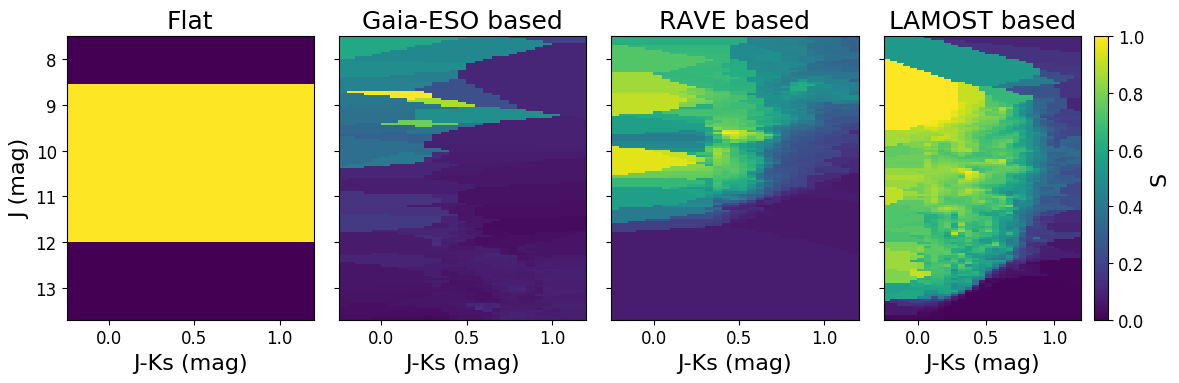}
    \caption{Input selection functions values ($S$) used in this work. This plot illustrates the function shape, hence the scale of $S$ is arbitrary here.}\label{fig:isf}
\end{figure*}

\subsection{Simulation set-up}
We test our method by applying it to simulated data, where the selection function is known.
We chose three different ``input'' selection functions (ISF). First function is defined analytically as:
\begin{equation}
  S(J, J-K) = \left\{
      \begin{array}{l}
  s_1, \textrm{if}\, 8^m.5 < J < 12^m \\
  0, \textrm{otherwise}
  \end{array}
   \right.\label{eq:s1},
\end{equation}
where $s_1$ is a constant. We refer to it as a ``constant'' ISF. Second and third functions are produced from estimates of the selection functions in three HEALPix cells for Gaia-ESO, RAVE and LAMOST surveys, multiplied by constants $s_2$, $s_3$ and $s_4$. In this way we simulate ``realistic'' selection functions. The shapes of the four ISFs used are shown in \autoref{fig:isf}.

For every ISF we use a number of stars to be sampled as ``observed'' $\Nobs$. This fixes the scaling constants $s_{1, 2, 3, 4}$, such that $\Nobs = \int \int S(J, J-K) \Nbg(J, J-K) dJ d(J-K)$. In each simulation we sample $\Nobs$ stars from the background $\Nbg(J, J-K)$ using the assumed selection function. The 2MASS background is taken from an arbitrary HEALPix cell. We have verified, that the result of the simulation depends much more on the parameters ($\Nobs$ and $\Ncri$) and adopted ISF than on the choice of HEALPix cell.
To increase the statistics, we run up to $n = 500$ simulations with different random samplings for each value of $\Nobs$. We then estimate the selection function $\tilde{S}$ (using \autoref{eq:selection}) for each simulated star and compare it with the ``true'' value $S$. We are interested in several parameters that will indicate the accuracy and the precision of the estimate. The first parameter we measure is the fractional uncertainty $U = \left\langle\uncert / \tilde{S}\right\rangle$, where $\uncert$ is the uncertainty of $\tilde{S}$. Angle brackets stand for mean or median taken over all stars in all $n$ simulations. The fractional uncertainty is a measure of the precision of the method, while the accuracy is measured by the relative bias:
\begin{equation}
 B = \left\langle\frac{\tilde{S} - S}{\uncert} \right\rangle. \label{eq:bias}
\end{equation}
It is also important to know if our uncertainty is realistic. This can be verified by testing the standard deviation of the relative difference:
\begin{equation}
  D = \sqrt{\frac{\sum \left(\frac{\tilde{S} - S}{\uncert} - B\right)^2 }{n \Nobs}}. \label{eq:stddev}
\end{equation}
In case the distribution of the difference between estimated and true values ($\tilde{S} - S$) is normally distributed, $\uncert$ should be close to the standard deviation of $\tilde{S} - S$. Hence, $D$ should be close to unity when the uncertainties are realistic. If it is lower than unity than we can suspect that the uncertainty is overestimated. Likewise, values larger than unity indicate that the uncertainty is likely underestimated. This approach works best, when the difference between $\tilde{S}$ and $S$ is normally distributed. In our simulations this is however not the case. Thus we use along with $D$ also a relative median absolute deviation (MAD):
\begin{equation}
 D_{MAD} = 1.48 \cdot \textrm{median}\left( \frac{|\tilde{S} - S|}{\uncert} - B \right),\label{eq:mad}
\end{equation}
with a constant ($1.48$) used to ensure that $D_{MAD} = D$ if $\tilde{S} - S$ is normally distributed.

We test different methods to estimate the selection function and its uncertainty (see \autoref{sec:uncertainty}) in each case:
\begin{itemize}
    \item Histogram binning (\autoref{sec:hess_binning});
    \item Median division binning (\autoref{sec:median_binning});
    \item Median division binning, with shrinking and interpolation (\autoref{sec:improvements});
\end{itemize}
For the median division binning, we also applied shrinking and interpolation separately -- this was done to study the effect of each of them on the precision and accuracy.

When the median division binning is used, a free parameter appears, namely, the minimal number of points in each cell: $\Ncri$ (see \autoref{sec:median_binning}). We run a set of simulations with different values of $\Ncri$ to explore the influence of this parameter on the results.

\subsection{Results of tests on simulated data}
Here we discuss how different methods and simulation parameters affect the precision and accuracy of the selection function estimation. The results are summarised in \autoref{fig:methods} where we explore how the precision and accuracy vary with $\Nobs$; and in \autoref{fig:methods2} where the effect of varying $\Ncri$ is shown.  We show here results for the ``constant`` and RAVE-based ISF. Results for Gaia-ESO-based and LAMOST-based ISF are qualitatively very similar to those for RAVE-based ISF.

\subsubsection{Sensitivity to $\Nobs$}
\begin{figure}
    \myimagesmall{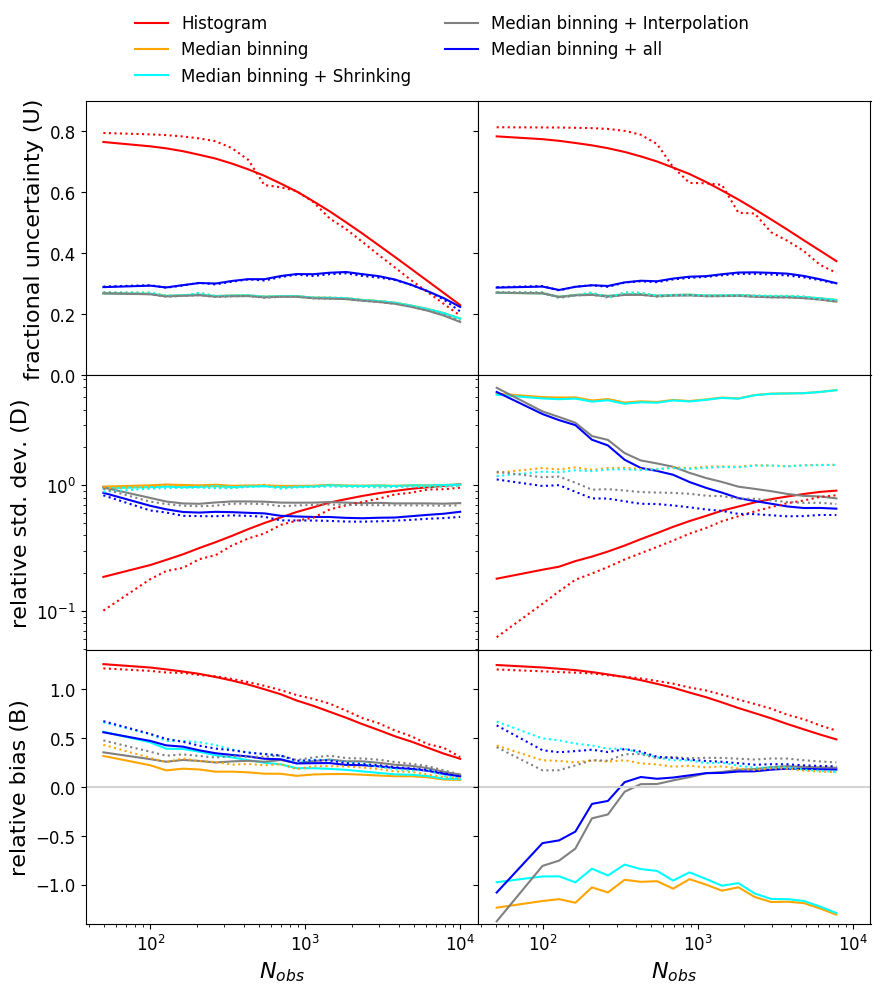}
    \caption{Comparison of the performance of different methods as a function of $\Nobs$, applied for ``constant'' input selection function (left column) and RAVE-based ISF (right column). Results are shown as a function of the number of ``observed'' stars $\Nobs$. Top row: fractional uncertainty; middle row: relative standard deviation and relative median absolute deviation (see \autoref{eq:stddev} and \ref{eq:mad}); bottom row: relative bias $B$ (see \autoref{eq:bias}). 
    Solid lines are for mean values, dotted lines are for median values. $\Ncri = 10$ is used here.}\label{fig:methods}
\end{figure}

\autoref{fig:methods} illustrates how results for simulated data depend on the method and on the number of ``observed'' stars $\Nobs$.
We expect results to improve with increasing $\Nobs$, as $\Nfg$ in Equations \ref{eq:selection} and \ref{eq:uncertainty} increases with $\Nobs$, which in turn causes the fractional uncertainty to decrease. At the same time, larger values of $\Nfg$ reduce biases, as the colour-magnitude diagram becomes more populated with increasing $\Nfg$, which increases the number of populated histogram bins and reduces the cell sizes for median division binning. This allows us to improve the accuracy of the selection function estimate.

Methods based on histograms show the largest fractional uncertainties $U$, especially for low $\Nobs$. This is caused by the fact that at low values of $\Nobs$ there are many underpopulated bins with one or few foreground sources. For such bins the uncertainty is large (up to 100\%), which has an impact on the mean fractional uncertainty.
Large uncertainty leads to low relative standard deviation $D$, when histograms are used to estimate the selection function, especially for low values of $\Nobs$ (see middle panels of \autoref{fig:methods}). The large bias $B$ is caused by the fact that the selection function is being measured only at the ``observed'' stars locations, which causes a positive bias (see \autoref{sec:hess_binning} and bottom panels of \autoref{fig:methods}).
As $\Nobs$ increases, the fractional uncertainty and relative bias decrease, though remain highest among all methods. The relative standard deviation approaches unity, as expected.
Note, that the histogram method gives similar results for both presented ISFs. Mean and median values of $U$, $D$ and $B$ are also very similar.
 
In our simulations, median binning with $\Ncri = 10$ is used, and fractional uncertainties $U$ are smaller by about $10^{-1/2} \approx 0.32$ (see \autoref{eq:selection_approx} and \ref{eq:uncertainty_approx}), compared to uncertainties produced by histogram method. This is caused by a larger number of ``observed'' stars $\Nfg$ in each colour-magnitude cell compared to the histogram method, which causes the uncertainty to decrease (see \autoref{eq:uncertainty} and \ref{eq:uncertainty_approx}). 
As $\Nobs$ increases, more and more median binning cells reach the limit of the smallest possible cell size (which is the size of the colour-magnitude histogram bin). For such cells the number of the ``observed'' stars $\Nfg$ will be larger than $\Ncri$, which will cause the uncertainty to decrease further. At the same time the fractional uncertainty increases if the variations in the back- and foreground source counts within the cell are taken into account following \autoref{sec:bg_variation}. Both effects are relatively small and, as we will show below (in \autoref{sec:ncritical}), the value of $\Ncri$ has a much larger effect on the fractional uncertainty than the value of $\Nobs$ and the correction for variation in the back- and foreground source counts.

For the median binning, the relative standard deviation $D$ and the relative bias $B$ behaviour as a function of $\Nobs$ differs for the ``constant'' and RAVE-based ISFs.
These differences are more prominent for mean values than for median ones. This is a consequence of the asymmetrical distribution of relative difference $\frac{\tilde{S} - S}{\uncert}$ for simulated stars. This asymmetry arises from the fact that the precision of the estimate depends on the number of fore- and background sources and the size of the cell. Higher numbers of sources give larger statistics and smaller cells thus reducing the effect of fore- and background source density variations within the cell (see \autoref{sec:bg_variation}). Therefore, the precision will vary from cell to cell, and the distribution of relative difference will thus be non-Gaussian, causing the differences between mean and median values.

For the ``constant'' ISF, the relative standard deviation $D$ is close to unity, and is almost independent of $\Nobs$. When the interpolation is introduced (see \autoref{sec:interpolation}), $D$ decreases to values about $0.7$, which indicates that the uncertainty is overestimated. This is a consequence of the interpolation -- it tends to improve the precision and accuracy \citep{Interpolation}. The relative bias $B$ is positive, however lower than that for the histogram method. The reason is likely the same as for the histogram method -- we are positively biased, because the selection function is measured at locations of ``observed'' stars. This is  more important in the outer regions of the colour-magnitude diagram, where the number of background and ``observed'' stars is low. Shrinking (see \autoref{sec:shrinking}) enhances this effect, further increasing the relative bias $B$. Shrinking has little effect on fractional uncertainty $U$ and relative standard deviation $D$, and it's effect on the relative bias $B$ seems to be small in most cases, except small $\Nobs$. This is because shrinking affects only outer cells in the colour-magnitude diagram, and thus the total fraction of affected stars is small. However, for this small fraction of cells the effect can be substantial, eliminating very small values of the selection function.

For the RAVE-based ISF, mean and median values of the relative standard deviation $D$ and relative bias $B$ are different, because the variations of the ISF over colour-magnitude space produce a large and asymmetric spread in differences between true and estimated values of the selection function. 
 Median values of $D$ and $B$ change little with $\Nobs$, while mean values vary substantially.

If no interpolation is applied, large cells produced by median binning cannot properly trace small-scale variations of the ISF. This results in a substantially underestimated uncertainty and thus mean relative standard deviation values are much larger than unity. The inability to reproduce ISF variations without interpolation also leads to a large negative mean bias (around $-1$), though one has to keep in mind that the underestimated value of the uncertainty leads to overestimated absolute value of the mean relative bias.

The use of interpolation for the RAVE-based ISF reduces both relative standard deviation $D$ and relative bias $B$. As $\Nobs$ increases, the difference between mean and median values of $D$ and $B$ decreases, which means that the distribution of differences between true and estimated values of the selection function becomes more symmetric. 
There is an important critical point where the mean relative bias becomes zero (at around $\Nobs = 300$), which we consider as an optimal value for $\Ncri = 10$. For this value of $\Nobs$ the mean relative standard deviation is also close to unity, which indicates that the uncertainty value is correctly estimated.

\subsubsection{Sensitivity to $\Ncri$}\label{sec:ncritical}
\begin{figure}
    \myimagesmall{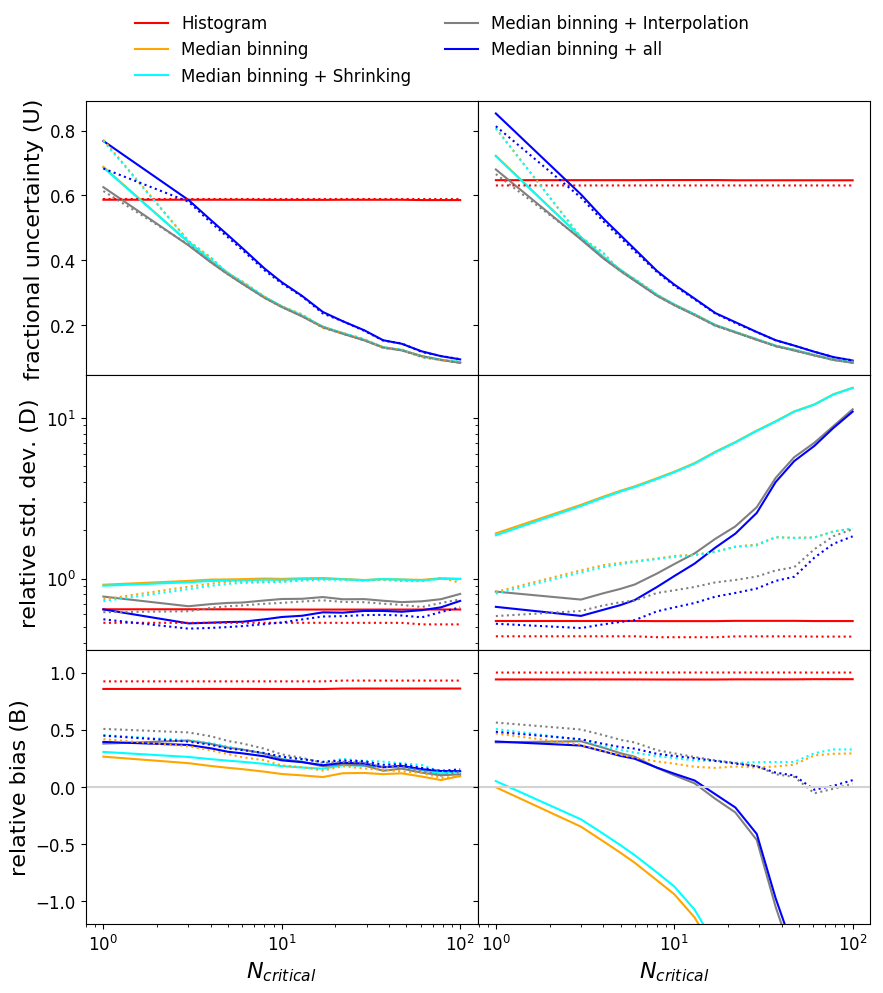}
    \caption{Same as \autoref{fig:methods}, now as a function of $\Ncri$. $\Nobs = 1000$ is used here.}\label{fig:methods2}
\end{figure}

\autoref{fig:methods2} illustrates how results from the simulations depend on the method and the minimal number of ``observed'' stars per colour-magnitude cell $\Ncri$. For this set of tests we set $\Nobs = 1000$. By definition, results for the histogram-based estimate do not depend on $\Ncri$. As we increase $\Ncri$, the fractional uncertainty of the estimate made with the median division binning method decreases approximately as $\Ncri^{-1/2}$, as expected, as $\Ncri$ is the lower limit for the value of $\Nfg$ in \autoref{eq:selection_approx} and \ref{eq:uncertainty_approx}. 
When background variations within the cell are taken into account, we find that the fractional uncertainty increases by about 15 percent (see top panel in \autoref{fig:methods2}) for both ``constant'' and RAVE ISFs. 

For the ``constant'' ISF, the relative standard deviation varies little with $\Ncri$ and the relative bias decreases slowly with $\Ncri$. 

For the RAVE-based ISF the trends are very different. The mean relative standard deviation is larger than unity for the median binning methods without interpolation and reaches values over ten for high $\Ncri$. This is caused by the fact that median binning tends to produce larger cells in the colour-magnitude space for larger values of $\Ncri$, within which the variation in the selection function is high and cannot be properly taken into account. Interpolation reduces the effect, though does not remove it completely.
Note that for $\Ncri = 100$ and the chosen value of $\Nobs = 1000$ we get in the best case 10 cells in the colour-magnitude diagram. With this low number of cells and hence a low number of interpolation points, it is impossible to properly reconstruct a two dimensional selection function of a complex shape. Median values of the relative standard deviation $D$ are nonetheless between 0.5 and 2, indicating that high mean values of $D$ are caused by a few large offsets, while for the majority of cases the uncertainty is estimated with reasonable quality. 

The mean relative bias for the RAVE-based ISF varies a lot when median binning is used, and goes from positive to negative values as $\Ncri$ increases. This is again caused by large cells produced by the median binning for large values of $\Ncri$. Our inability to reconstruct rapid variations of the selection function within a cell leads to large biases. This happens only in a fraction of cells where the selection function variations are large, and thus only the mean relative bias is affected, while the median relative bias remains between 0 and 0.5 and varies only slowly with $\Ncri$. Without interpolation, only for $\Ncri = 1$ the mean bias is close to zero. 
Interpolation improves the mean relative bias value for $\Ncri \approx 10$. Still, for large $\Ncri$ the mean bias decreases to below $-1$. The optimal point where the mean relative bias turns zero is $\Ncri \approx 15$, with the mean relative standard deviation being close to unity at this point, which indicates that the uncertainty value is correctly estimated.

\subsubsection{Selecting optimal parameters}

\begin{figure*}
    \begin{center}
        \myimage{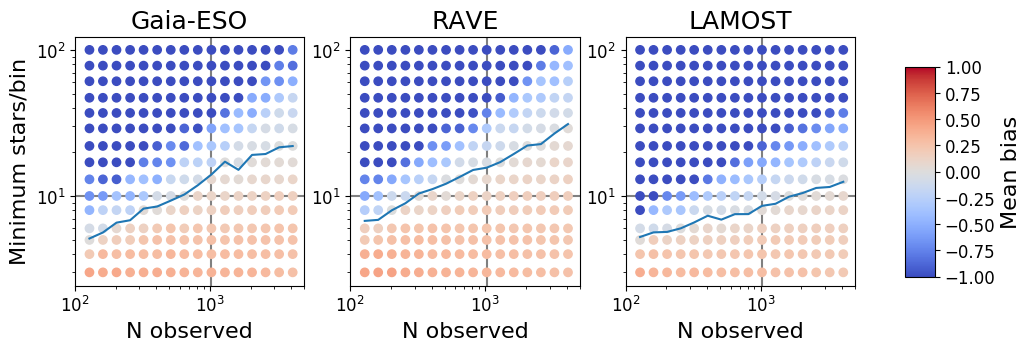}
    \end{center}
\caption{Mean relative bias $B$ as a function of the number of ``observed'' stars in simulations $\Nobs$ and the minimum number of stars per median division binning cell $\Ncri$. Grey lines are at $\Ncri = 10$ and $\Nobs = 1000$, used in \autoref{fig:methods} and \ref{fig:methods2}. Blue line indicates optimal values of $\Ncri$ for each $\Nobs$, where mean relative bias is zero (see \autoref{eq:optimal}).}\label{fig:twod}
\end{figure*}
We have shown in \autoref{sec:ncritical}, that the precision and accuracy of the selection function estimates depend on $\Ncri$. For a given number of ``observed'' stars $\Nobs$, there exists an optimal value of $\Ncri$ that minimizes the mean bias and at the same time produces a mean relative standard deviation $D$ (see \autoref{eq:stddev}) that is close to unity, which means that the  uncertainty is reliable. In order to find how the optimal $\Ncri$ depends on $\Nobs$, we run a set of tests for ISFs based on Gaia-ESO, RAVE and LAMOST data, varying both $\Ncri$ (between 3 and 100) and $\Nobs$ (between 100 and 4000). In \autoref{fig:twod} we show the mean relative bias as a function of $\Ncri$ and $\Nobs$, and indicate zero-bias lines. 
\begin{figure}
    \myimagesmall{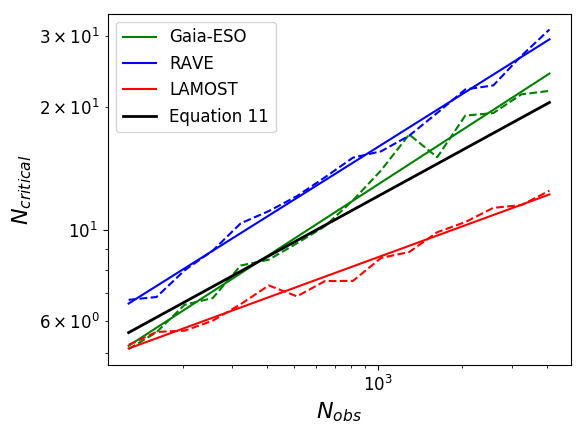}
    \caption{Optimal values of $\Ncri$ as functions of $\Nobs$ for three simulated ISFs (the dashed lines are the same lines as indicated in \autoref{fig:twod}), power-law fits (solid lines) and the accepted empirical relation (black line, see \autoref{eq:optimal})}\label{fig:fits}
\end{figure}
We fit zero-bias lines for Gaia-ESO- RAVE- and LAMOST-based ISFs with power-laws as $\Ncri = a \Nobs^b$. Zero-bias lines and power-law fits are shown in \autoref{fig:fits}. The three fitted functions differ by about a factor of two for a given $\Nobs$ which means that the optimal value of $\Ncri$ depends in addition to $\Nobs$ also on the shape of the selection function itself. Taking the mean parameters of the three fits, we obtain the following empirical relation, which we use in further study:
\begin{equation}
  \Ncri = int(0.91 * \Nobs^{0.37}) + 1. \label{eq:optimal}
\end{equation}
We note however, that the selection function estimate quality is not a very sensitive function of $\Ncri$, unless the value used is much higher than the optimal (see \autoref{fig:methods2}): the absolute value of the mean relative bias $B$ remains lower than 0.5 even if we vary $\Ncri$ within 50\% of the optimal value, with less variations for higher $\Nobs$.

\section{Data preparation}\label{sec:data}
We calculate the selection function for a large set of public surveys: APOGEE \citep[DR14][]{APOGEE}, Gaia-ESO \citep[DR2][]{GAIA_ESO}, GALAH \citep[DR2][]{GALAH_DR2}, LAMOST \citep[DR3][]{LAMOST}, LAMOST Galactic anti-centre project \citep{LAMOST_GAC_2}), RAVE \citep[DR5][]{RAVE_DR5}, RAVE-on \citep{RAVE_ON}) and SEGUE \citep{2009AJ....137.4377Y}.
For each survey, we process stars that satisfy the following criteria:
\begin{enumerate}
    \item have good quality 2MASS $J$ and $Ks$ photometry (corresponding Rflg value is 1, 2 or 3);
    \item for 2MASS magnitudes the following constraints hold: $1^m.4 < J < 15^m$ and $-1^m.5 < J-Ks < 2^m.5$ -- ranges of magnitudes and colours covered by 2MASS, excluding extremely red, blue and bright objects;
    \item repeated observations of the same star within the survey are excluded (thus we only consider one observation per star in each survey).    
\end{enumerate}
For some surveys, we had to apply more cuts or use additional information, as described below. 

\subsection{APOGEE treatment}
APOGEE survey is special in a sense that for some fields additional narrow-band photometry (using Washington M, T2 and DDO51 filters) was used to select giant stars over dwarfs \citep[see][]{2013AJ....146...81Z}. This cannot be accounted for, when only broad-band 2MASS photometry is used. To mitigate this problem, we added to our 2MASS background distribution a correction factor derived using the APOGEE photometric input catalogue. For each HEALPix cell, we calculated this correction factor as a ratio of stars in the APOGEE photometric input catalogue that satisfy the selection criteria for giant stars\footnote{\texttt{wash\_ddo51\_giant\_flag = 1} or extinction corrected colour $(J-Ks)_0 > 0.5$} to the total number of stars for this HEALPix cell in this input catalogue. This ratio was calculated as a function of $J-Ks$ colour and $J$ magnitude on the same grid that is used for the background distribution. 

\subsection{Gaia-ESO treatment}
Gaia-ESO aims, among other things, on studies of open clusters and the Milky Way in general. For fields used for open cluster studies, target allocation for spectroscopic observations favours possible cluster members. Therefore we cannot properly calculate the selection function for them, as the observed population is not a representative sample of the stellar population in a given field. Hence, we calculate the selection function only for Milky Way fields.

\subsection{Selection function estimation}
In the current work, we estimate selection function for each survey over the HEALPix grid of three different orders (3, 4 and 5, see \autoref{sec:healpix}). This is done to find a balance between the background variation over the HEALPix sky-cells (which typically increases as we increase the sky-cell size) and the number of spectroscopic sources observed (which decreases with decreasing sky-cell size). Only HEALPix sky-cells with at least 50 stars were processed, to ensure that there are enough stars for a reliable estimation of the selection function.

\section{Results}
We estimate the selection function values using the median binning method with all improvements as described in \autoref{sec:method} for all surveys mentioned in \autoref{sec:data}, and discuss the results here. The tables containing the selection function estimates are published at the UniDAM homepage\footnote{\url{http://www2.mps.mpg.de/homes/mints/selection.html}}. 
An example of the results table is shown in \autoref{tbl:sample}.
\begin{table*}
\begin{tabular}{crrrrrrc}
    \toprule
    id &  selection\_3 &  selection\_3\_err &  selection\_4 &  selection\_4\_err &  selection\_5 &  selection\_5\_err &  best\_order \\
    \midrule
  107548 &      0.00340 &          0.00143 &      0.01542 &          0.00646 &            - &                - &                     4 \\
  107549 &      0.00403 &          0.00155 &      0.01602 &          0.00613 &      0.03597 &          0.01136 &                     5 \\
  107551 &      0.01226 &          0.00476 &      0.05289 &          0.02022 &      0.07561 &          0.02492 &                     5 \\
  107554 &      0.00951 &          0.00387 &      0.04294 &          0.01720 &      0.07754 &          0.02284 &                     5 \\
  107557 &      0.00235 &          0.00105 &      0.01126 &          0.00500 &      0.01276 &          0.00572 &                     5 \\
  107561 &      0.00939 &          0.00376 &      0.04154 &          0.01638 &      0.11274 &          0.05250 &                     4 \\
  107562 &      0.00860 &          0.00352 &      0.03501 &          0.01423 &      0.06088 &          0.02832 &                     4 \\
  107564 &      0.00644 &          0.00271 &      0.02982 &          0.01237 &      0.06768 &          0.03117 &                     4 \\
  107568 &      0.00725 &          0.00296 &      0.03258 &          0.01314 &      0.06622 &          0.01872 &                     5 \\
  107571 &      0.00932 &          0.00372 &      0.04109 &          0.01615 &      0.07057 &          0.02147 &                     5 \\
  \bottomrule
\end{tabular}
\caption{An example of the result table. The first column is an ID of a star from the spectroscopic survey. Columns two to seven contain the value of the selection function and its uncertainty calculated using 3rd, 4th and 5th order HEALPix sky-cells. The last column indicates which of the HEALPix orders gives the lowest fractional uncertainty of the selection function.} \label{tbl:sample}
\end{table*}

\begin{figure*}
    \myimage{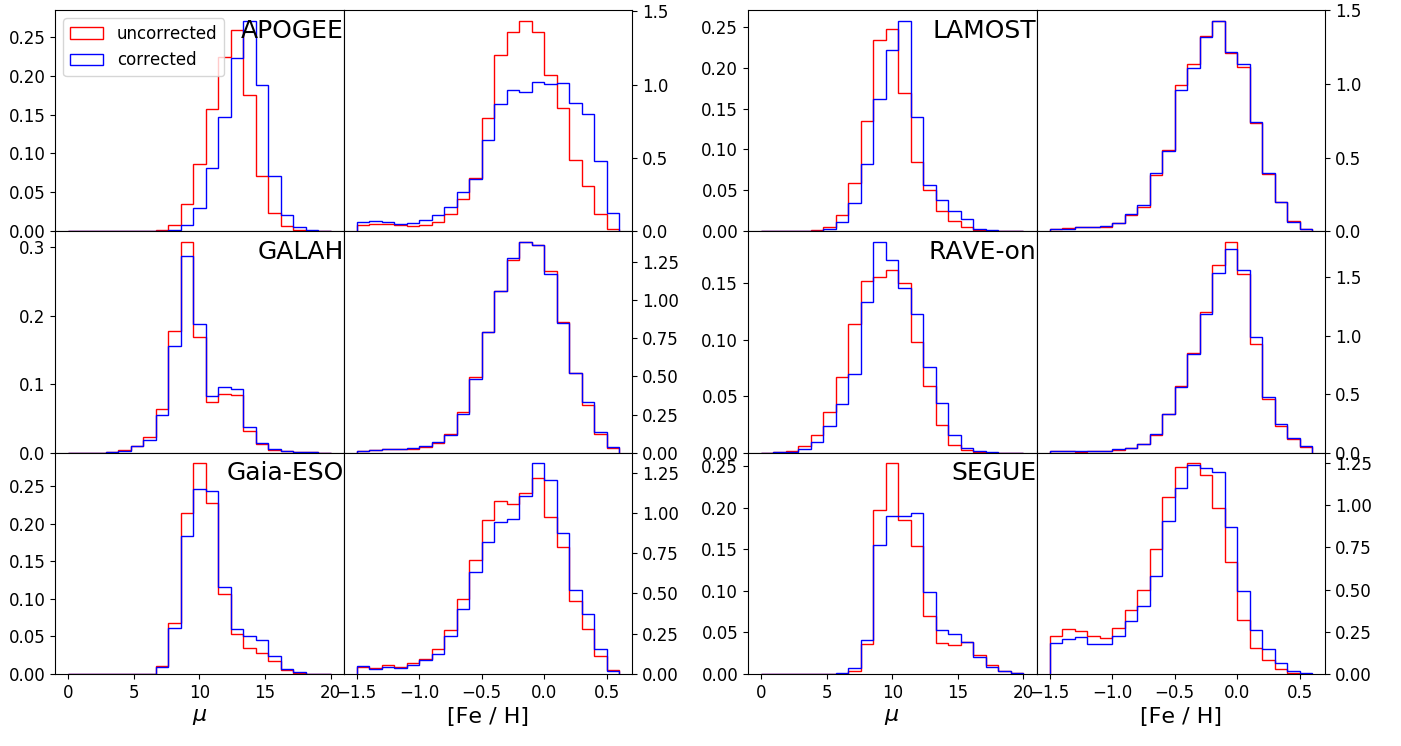}
	\caption{Normalized distributions in distance modulus (left) and metallicity (right) for APOGEE, GALAH, Gaia-ESO, LAMOST, RAVE-on and SEGUE surveys. Red histogram shows distribution not corrected for the selection effect, blue histogram -- with corrections applied.}\label{fig:results1}
\end{figure*}

In \autoref{fig:results1} we show the effect of the selection function on the metallicity and distance modulus distributions for several surveys. Distance moduli for stars were taken from UniDAM catalogue \citep{2017yCat..36040108M}. Stars that do not have an entry in the UniDAM catalogue were excluded from this analysis. Uncorrected distributions for value $x$ are made by counting stars in each bin:
\begin{equation}
  F_{\textrm{uncorrected}}(x) = \sum_{i: x - b < x_i < x + b} 1,
\end{equation}
where $b$ is the bin half-width.
The value of the selection function for a given star is by definition a fraction of the underlying population represented by that star. Thus the selection-corrected distribution is calculated like:
\begin{equation}
F_{\textrm{corrected}}(x) = \sum_{i: x - b < x_i < x + b} \frac{1}{S_i},
\end{equation}
where $S_i$ is the value of the selection function for the $i$-th star. Distributions shown in \autoref{fig:results1} are normalized so that $\int F(x)\,dx = 1$ in order to emphasize the change in the shape of the distributions rather than the change in scale. 

The uncertainties of $F_{\textrm{corrected}}(x)$ can be calculated by propagating the uncertainties of measurements of $S_i$:
\begin{equation}
  \sigma_F = \sqrt{\sum_{i: x - b < x_i < x + b} \left(\frac{\sigma_{S, i}}{S_i^2} \right)^2}.
\end{equation}

These uncertainties are small (on the order of 1-2 percent) for distributions shown in \autoref{fig:results1}, because of the large number of sources in each survey and hence are not displayed there.

The overall trend is that when the selection is corrected for both distance modulus and metallicity distributions we find a shift towards larger values. For the distance modulus distribution the change in the distribution is caused by the fact that more distant stars are systematically fainter, and for fainter stars the selection function is typically lower, which means that we observe a smaller fraction of more distant stars. The shift of the metallicity distributions is caused by the fact that surveys typically have nearly constant source density in their footprints, which means that we observe a smaller fraction of stars closer to the galactic plane, where stellar density is larger. At the same time, the average metallicity of thin disk stars in the galactic plane is higher than those high above the galactic plane. Therefore, we observe a smaller fraction of metal-rich stars, which is reflected in the selection function values and the selection-corrected distribution is shifted towards higher metallicities.

The effect of the selection function is more prominent for surveys that have a complex target allocation strategy and whose footprint is more patchy: APOGEE, Gaia-ESO and SEGUE. This is in line with findings of \citet{2016MNRAS.460.1131S} for Gaia-ESO and \citep{2017A&A...606A..97N} for both APOGEE and Gaia-ESO. For surveys with contiguous footprints, like GALAH, RAVE and LAMOST, the selection function has almost no effect, which confirms findings of \citet{2017MNRAS.468.3368W} and \citet{2018MNRAS.476.3278C}.

\begin{figure*}
    \myimageTwo{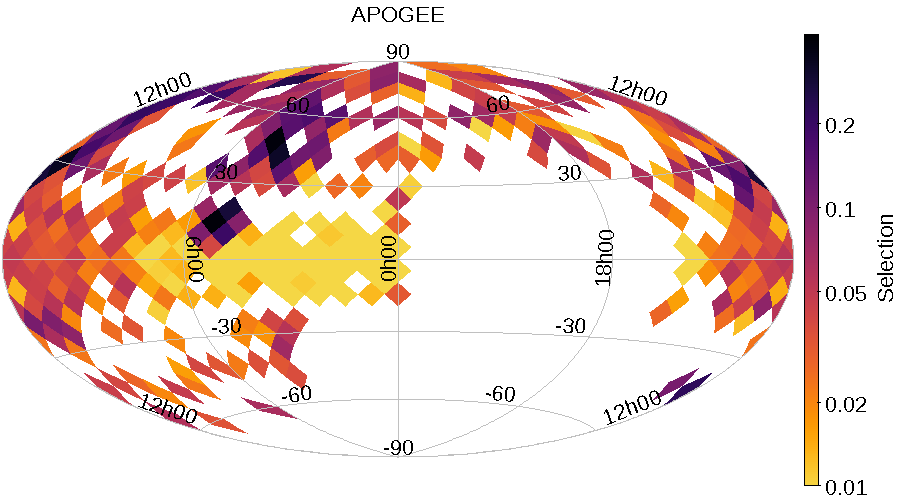}{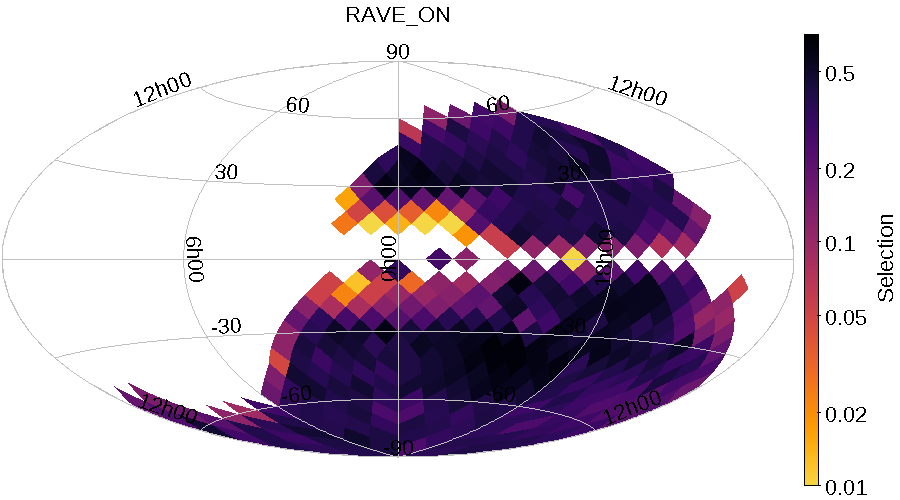}
    \caption{Median selection function value as a function of galactic coordinates for APOGEE (left) and RAVE-on (right) surveys. Note the different colour scales for the two plots.}\label{fig:sky}
\end{figure*}

\autoref{fig:sky} illustrates how the selection function varies across the sky for APOGEE and RAVE-on. The trend is that the selection function is lower towards the galactic plane and galactic centre. We emphasize that for APOGEE the selection function seems to be much more uneven than for RAVE-on, which will complicate any statistical analysis based on the spatial distribution of survey stars.

\section{Conclusion and discussion}
In this work we present a method that allows us to estimate the selection function for a general spectroscopic survey. Precision and accuracy of the method were verified with realistic simulations. These simulations also allowed to derive an empirical formula to estimate the only free parameter of the method, namely, the minimal number of observed stars ($\Ncri$, see \autoref{eq:optimal}) per colour-magnitude cell. 

Our estimates can be readily used to correct for the selection effects in galactic archaeology studies. If a subset of the survey for which the selection values are published is used, two possibilities arise. We will explain them with two examples. 

If the subset is constructed using some parameter $p$ that is not directly connected to the properties of the stellar population, than one can build a new selection function as $S' = S(J, J-Ks) \times F(p)$, where $F(p)$ is the subset selection function with respect to the complete spectroscopic survey. This is possible, because the subset remains a representative part of the full underlying population. As an example of such parameter $p$ we can name the signal-to-noise ratio (SNR) of the spectra. 

Another case is when the subset is build using some property of the population. An example of such subsets can be high-velocity stars or metal-poor stars. In that case the subset is no longer a representative part of the underlying population. This means that with that subset we can only study the part of the full stellar population that it represents. The selection function of the complete survey can still be used to correct for the selection effects in the subset. Let us consider for example a cell in the colour-magnitude diagram for some field on the sky, for which 100 stars were observed spectroscopically out of 1000 stars in the photometric catalogue (2MASS) for that cell. This gives a selection function value $S=0.1$, or 10 photometric stars per 1 spectroscopic. If five of the spectroscopically observed stars are, for example, extremely metal poor, we cannot assume that the selection function of them is $S = 5 / 1000$, as this would imply that all photometric stars in that cell are in fact metal poor, which is not true. Though we can use $S=0.1$ to estimate that $n = 5 / S = 50$ stars are metal poor out of 1000 for that cell.

We produce and make public the estimates of the selection function values and their uncertainties for a set of public spectroscopic surveys. The tool to produce such estimates will be made available on the MPS github page\footnote{\url{https://github.molgen.mpg.de/mints/sage_selection_public}}.

For some surveys, like LAMOST, GALAH and RAVE, the effect of the selection function is negligible, at least when the distributions of distances and metallicities are considered. For other surveys the effect of the selection function is visible in the distributions of distances and metallicities. This is the case for Gaia-ESO, SEGUE and to a larger extent for APOGEE, where ignoring the selection effect might produce a substantial bias. 

Values of the selection function calculated in this work can be used to estimate the fraction of stars with given properties that were observed as compared to those available for observations in the footprint of a given survey. This is an essential step towards calculating the fraction of all stars in the footprint of the survey that were observed. 

\section*{Acknowledgements}
The research leading to the presented results has received funding from the European Research Council under the European Community's Seventh Framework Programme (FP7/2007- 2013)/ERC grant agreement (No 338251, StellarAges). 

This research has made use of the VizieR catalogue access tool, CDS, Strasbourg, France. This research made use of Astropy, a community-developed core Python package for Astronomy \citep{2013A&A...558A..33A} This research made use of matplotlib, a Python library for publication quality graphics \citep{Hunter:2007} This research made use of TOPCAT, an interactive graphical viewer and editor for tabular data \citep{2005ASPC..347...29T}  This publication makes use of data products from the Two Micron All Sky Survey, which is a joint project of the University of Massachusetts and the Infrared Processing and Analysis Center/California Institute of Technology, funded by the National Aeronautics and Space Administration and the National Science Foundation. Funding for SDSS-III has been provided by the Alfred P. Sloan Foundation, the Participating Institutions, the National Science Foundation, and the U.S. Department of Energy Office of Science. The SDSS-III web site is http://www.sdss3.org/. SDSS-III is managed by the Astrophysical Research Consortium for the Participating Institutions of the SDSS-III Collaboration including the University of Arizona, the Brazilian Participation Group, Brookhaven National Laboratory, University of Cambridge, Carnegie Mellon University, University of Florida, the French Participation Group, the German Participation Group, Harvard University, the Instituto de Astrofisica de Canarias, the Michigan State/Notre Dame/JINA Participation Group, Johns Hopkins University, Lawrence Berkeley National Laboratory, Max Planck Institute for Astrophysics, Max Planck Institute for Extraterrestrial Physics, New Mexico State University, New York University, Ohio State University, Pennsylvania State University, University of Portsmouth, Princeton University, the Spanish Participation Group, University of Tokyo, University of Utah, Vanderbilt University, University of Virginia, University of Washington, and Yale University.  Guoshoujing Telescope (the Large Sky Area Multi-Object Fiber Spectroscopic Telescope LAMOST) is a National Major Scientific Project built by the Chinese Academy of Sciences. Funding for the project has been provided by the National Development and Reform Commission. LAMOST is operated and managed by the National Astronomical Observatories, Chinese Academy of Sciences. Funding for RAVE has been provided by: the Australian Astronomical Observatory; the Leibniz-Institut fuer Astrophysik Potsdam (AIP); the Australian National University; the Australian Research Council; the French National Research Agency; the German Research Foundation (SPP 1177 and SFB 881); the European Research Council (ERC-StG 240271 Galactica); the Istituto Nazionale di Astrofisica at Padova; The Johns Hopkins University; the National Science Foundation of the USA (AST-0908326); the W. M. Keck foundation; the Macquarie University; the Netherlands Research School for Astronomy; the Natural Sciences and Engineering Research Council of Canada; the Slovenian Research Agency; the Swiss National Science Foundation; the Science \& Technology Facilities Council of the UK; Opticon; Strasbourg Observatory; and the Universities of Groningen, Heidelberg and Sydney. The RAVE web site is at \url{https://www.rave-survey.org}.  Based on data products from observations made with ESO Telescopes at the La Silla Paranal Observatory under programme ID 188.B-3002. These data products have been processed by the Cam-bridge Astronomy Survey Unit (CASU) at the Institute of Astronomy, University of Cambridge, and by the FLAMES/UVES reduction team at INAF/Osservatorio Astrofisico di Arcetri. These data have been obtained from the Gaia-ESO Survey Data Archive, prepared and hosted by the Wide Field Astronomy Unit, Institute for Astronomy, University of Edinburgh, which is funded by the UK Science and Technology Facilities Council. 

\bibliographystyle{aa.bst}
\bibliography{selection}

\end{document}